\documentclass[final,5p,times,twocolumn,authoryear]{elsarticle}

\usepackage{amssymb}
\usepackage[T1]{fontenc}
\usepackage[utf8]{inputenc}
\usepackage{amsmath}
\usepackage{graphicx}
\usepackage[
  citecolor=blue,
  colorlinks,
  linkcolor=blue,
  urlcolor=blue,
]{hyperref}
\usepackage[dvipsnames]{xcolor}
\usepackage[makeroom]{cancel} 

\renewcommand{\b}{\hat{\bb{b}}}

\newcommand{\be}{\begin{eqnarray}}
\newcommand{\en}{\end{eqnarray}}
\newcommand{\pa}{\partial}
\newcommand{\f}{\frac}

\newcommand\bb[1]{\mbox{\boldmath{$#1$}}}
 
\newcommand\bcdot{\bb{\cdot}}
\newcommand\btimes{\bb{\times}}

\begin{document}

\begin{frontmatter}

\author[iitk]{H. ~Gupta\corref{cor1}}
\ead{hiugupta@iitk.ac.in}

\author[iitk]{S.K. Rathor}
\ead{skrathor@iitk.ac.in}

\author[nbi]{M.E. Pessah}
\ead{mpessah@nbi.dk}

\author[iitk,iitkm]{S. Chakraborty}
\ead{sagarc@iitk.ac.in}

\cortext[cor1]{Corresponding author}

\address[iitk]{Department of Physics, Indian Institute of Technology Kanpur, 
U.~P.-208016, India}
\address[nbi]{Niels Bohr International Academy, Niels Bohr Institute, 2100,
Copenhagen \O, Denmark}
\address[iitkm]{Mechanics \& Applied Mathematics Group,
Indian Institute of Technology Kanpur, U.P.-208016, India}

\title{Stability Analysis of Convection in the Intracluster Medium}

\begin{abstract} 
We use the machinery usually employed for studying the onset of Rayleigh--B\'enard
convection in hydro- and magnetohydro-dynamic settings to address
the onset of convection induced by the magnetothermal instability and the
heat-flux-buoyancy-driven-instability in the weakly-collisional 
magnetized plasma permeating the intracluster medium.  
Since most of the related numerical simulations consider the plasma being bounded
between two `plates' on which boundary conditions are specified, our
strategy provides a framework that could enable a more direct connection 
between analytical and numerical studies.
We derive the conditions for the onset of these instabilities considering 
the effects of induced magnetic tension resulting from a finite plasma beta.
We provide expressions for the Rayleigh number in terms of the 
wave vector associated with a given mode, which allow
us to characterize the modes that are first to become unstable.
For both the heat-flux-buoyancy-driven-instability and the magnetothermal instability, oscillatory marginal stable states are possible.
\begin{description}
\item[PACS numbers]{98.65.Hb, 44.25.+f}
\end{description}
\end{abstract}


\end{frontmatter}


\section{Introduction} 
\label{sec:Introduction}

Convection, i.e., the motions induced within a fluid by the tendency
of hotter, less dense material to rise, and colder, denser material to
sink under the influence of gravity, is a ubiquitous phenomenon in
nature.  These motions, and the 
ensuing transfer of heat, can have
important implications for a wide variety of systems, see e.g., \citet{1999WS.book.....}, 
ranging from laboratory settings to the Earth and the oceans, from planetary to stellar
atmospheres, and from accretion disks
\citep{1996ApJ...464..364S, 2010MNRAS.404L..64L, 2012ApJ...761..116B, 
2013ApJ...770..100G}  to the intracluster medium (ICM) permeating galaxy clusters
\citep{2001ApJ...562..909B, 2008ApJ...673..758Q, 2009ApJ...703...96P, 2009ApJ...704..211B, 
2011MNRAS.413.1295M, 2012ApJ...754..122K}.

The inherent nonlinearity of the governing equations, together with the complex dynamical
boundaries present in nature, has motivated the study of convection in
idealized settings where the fluid is
confined between two parallel horizontal plates and is heated from below.  When this setup
leads to convective motions, this is termed Rayleigh-B\'enard
convection (RBC).
The stability of the equilibrium state and the flow dynamics in RBC
are determined by a non-dimensional parameter viz., the Rayleigh
number $R$, which is a measure of the strength of the destabilizing 
buoyancy force relative to the stabilising viscous force in the fluid. 
When the Rayleigh number for a given fluid is below a
critical number, then heat transfer occurs primarily via conduction;
when this critical number is exceeded, heat transfer is primarily via
convection.  The onset of the instability and the critical value of
$R$ can be understood by means of a linear stability analysis
\citep{1981HH.book.....}.

The rich nonlinear phenomena (e.g., pattern formation, route to chaos,
turbulence, etc.) ensuing in such a convective system can also be
analytically investigated in the weakly nonlinear limit
\citep{1987CCF.book....., 2009PFDS.book.....}.  There is a large body
of literature on flow reversals, pattern formation and evolution in RBC encompassing
both experiments \citep{1993PRL...71..2026, 1996PRL...76..756} as well
as nonlinear two-dimensional (2D) \citep{PhysRevLett.110.114503} and
three dimensional (3D) simulations \citep*{2003PRE...67..046313}.

The study of RBC has benefited the understanding of convection in a
wide variety of systems in nature, for instance, in the Earth's outer
core \citep*{{1994PhyEarthPlanetInter...82..235}}, mantle
\citep*{1974JFM...62..465}, atmosphere \citep*{2001JClim...14..4495},
and oceans \citep{1999RevGeophy...37..1}, as well as in Sun spots
\citep*{2003ApJ...588..1183}, and in metal production processes
\citep*{1988NHT...13..297}.  The framework employed to study RBC has
been generalized by considering the presence of magnetic fields and
even incorporating the effects of rotation, a combination prevalent in
astrophysical fluids. This approach has shed light into the generation
and reversal of the Earth magnetic field \citep{1995Nature...377..203}
and the internal dynamics of the Sun \citep{1996JFM...306..325,
2003ApJ...588..1183}.

In all of the cases in which conducting media have been considered,
the plasma has been assumed to behave as a magnetized fluid,
as described in the magnetohydrodynamic (MHD) approximation.
There are situations of astrophysical interest, however, in which the plasma
is only weakly-collisional, i.e., the mean free path for particles to 
interact is much larger than the Larmor radius. This is the case for
the dilute ICM permeating galaxy clusters, in which
transport properties are anisotropic with respect to the direction of 
the magnetic field.

The aim of this letter is to build upon the machinery employed to study
RBC in hydro- and magnetohydro-dynamic scenarios in order to address
the onset of convection in the weakly-collisional magnetized plasma in
galaxy clusters.

\subsection{Instabilities in the Weakly-Collisional ICM}

The ICM is a weakly collisional and high-beta plasma (see e.g.,
\citep{2002ARA&A..40..319C, 2006physrep...427..1}), in which the
transport of heat, transport of momentum and diffusion of ions is anisotropic due
to the presence of magnetic field.  Linear stability analysis has
shown that the ICM is dynamically unstable, to the so-called
magnetothermal instability, MTI, \citep{2000ApJ...534..420B, 2001ApJ...562..909B} and the
heat-flux-driven buoyancy instability, HBI \citep{2008ApJ...673..758Q}. 
The MTI sets in when the
temperature gradient decreases outwards and the magnetic field lines
are perpendicular to the direction of gravity, whereas the HBI is
excited when the temperature gradient increases outwards and the
magnetic field lines are parallel to the gravitational field.  The
original studies of these instabilities have been generalized to
explore the effects of viscous anisotropy \citep{2010POP...042117..17,
2011MNRAS.417..602K} and semi-global settings
\citep{2012MNRAS.423.1964L}. More recently, 
\citet{2013ApJ...764..13, 2015ApJ...813...22B}
analyzed the stability of the ICM generalizing previous work by
considering the effects of concentration gradients that could be
present in the ICM if the sedimentation of Helium is effective
\citep{2004MNRAS.349L..13C, 2009ApJ...693..839P, 2010MNRAS.401.1360S}.

Researchers have carried out nonlinear numerical studies of the MTI \citep{2005ApJ...633..334P, 2007ApJ...664..135P, 2008ApJ...688..905P,2011MNRAS.413.1295M} and the HBI \citep{2008ApJ...677L...9P,2009ApJ...703...96P, 2010ApJ...712L.194P,2011MNRAS.413.1295M,2012ApJ...754..122K} in connection with the `cooling
flow problem' in cool core galaxy clusters.  The effects of shear flow
(and thus, Kelvin-Helmholtz instability) on the stability condition
for MTI is explored in \citet{2011POP...022110..18}.  Recently,
\citet{2014APJ...792..1} 
performed linear stability analysis on weakly magnetized, rotating
plasma in both collisional and collisionless environments, leading to
more complete picture of ICM.

\subsection{Advantages of the Rayleigh--B\'enard Approach} 
\label{sec:benefitsI}

There are a number of advantages that follow from employing the
machinery developed for RBC to the study of the MTI and HBI.
This approach allows us to shed light into many aspects of the MTI and
the HBI, which are thought to play a role in the dynamics of the
intracluster medium (ICM). For instance,

\begin{enumerate} 

\item This framework provides a good platform to several connections with
numerical simulations because the boundary conditions (BCs) usually
adopted resemble the ones employed in RBC.

\item The results obtained can help us identify the critical Rayleigh number 
for the onset of the MTI and the HBI.

\item The formalism allows us to account for the effects of magnetic
tension on the stability criterion for both the MTI and the HBI. This
approach could be useful in order to assess the effects of magnetic
tension on the unstable growing modes found to feed off composition
gradients in a inhomogeneous intracluster medium
\citep{2013ApJ...764..13, 2015ApJ...813...22B}

\item The analysis could enable a low dimensional model like the
Lorenz model for RBC \citep*{1963JASci...20..571,
2006ChaosSolutionsFractals...28..571} and magnetic RBC
\citep{2003TAMech...30..29}, which could give further insights
into the chaotic (turbulent) state of the ICM.

\end{enumerate}

\section{The Rayleigh--B\'enard Framework} \label{sec:RBA}
\begin{figure}
 \centering
     \includegraphics[scale=0.95]{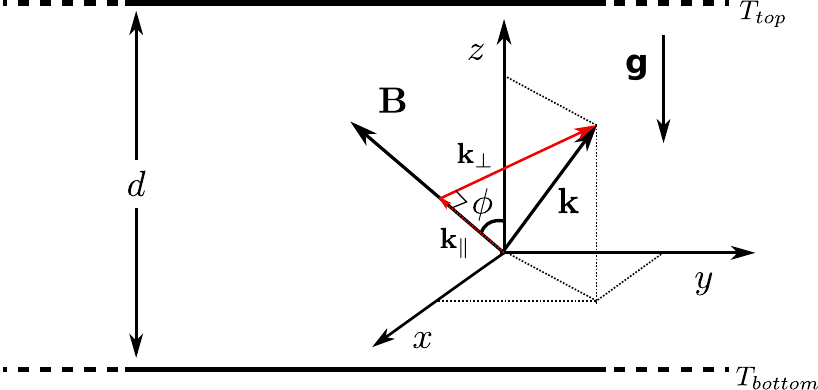}
     \caption{Schematic representation of the model geometry employed to study 
     Rayleigh-B\'enard convection (RBC). The dilute, weakly-collisional, 
     magnetized plasma  is held between two conducting horizontal plates of 
     infinite extent, separated by a distance $d$. 
     The plates are maintained at two different constant 
     temperatures as indicated. Gravity is along the negative $z$ axis. The 
     symbols $\perp$ and $\parallel$ represent the directions perpendicular 
     and parallel to the magnetic field, which lies in the $x-z$ plane.}
     \label{fig:geometry}
\end{figure}

Let us consider a weakly-collisional plasma at rest confined between
two horizontal parallel plates of infinite extent, as it is shown in
Fig. \ref{fig:geometry}.  The vertical separation between the plates
is $d$ and the acceleration due to gravity $\bb{g}$ is acting
vertically downwards.  The bottom and the top boundaries are held at
two different constant temperatures $T_{\rm bottom}$ and $T_{\rm
top}$, respectively.  This sets up a constant background temperature
gradient in the confined plasma.  There is also an externally imposed
uniform magnetic field $\bb{B}$ lying on the $x-z$ plane and acting on
the system under study.

\subsection{Governing Equations}
The equations of motion describing the dynamics of this system are given by
\be &&\f{\pa \rho}{\pa t}+\bb{\nabla}\bcdot(\rho \bb{v})=0 \,,
\label{eq:rho}\\
&&\f{d \bb{v}}{dt} =-\f{1}{\rho}\bb{\nabla} \bcdot \left(\mathsf{P} + 
\f{\bb{B}^2}{8\pi}\mathsf{I} - \f{{B^2}}{4\pi}\hat{\bb{b}}\hat{\bb{b}}\right) + \bb{g} \,, \,\,\,\,
\label{eq:v}\\
&&\f{\pa \bb{B}}{\pa t}=\bb{\nabla}\btimes(\bb{v}\btimes\bb{B})+ \eta \nabla^2\bb{B}\,,
\label{eq:b2}\\
&& \rho T\f{ds}{dt}=
(p_\bot-p_\parallel)\f{d}{dt}\ln\f{B}{\rho^{\gamma-1}} 
- \bb{\nabla} \bcdot \bb{Q}_{\rm s} 
 \,.
\label{eq:S} 
\en
Here, the Lagrangian and Eulerian derivatives are related via
$d/dt \equiv \pa/\pa t + \bb{v} \bcdot \nabla$, where 
$\bb{v}$ is the fluid velocity. 
The symbols $\rho$, $T$, $s$, $\gamma$ and $\eta$ stand for the fluid 
density, temperature (assumed to be the same for ions and electrons), 
specific entropy, adiabatic index and electrical resistivity (also called magnetic diffusivity).

The weakly collisional character of the plasma renders its
physical properties anisotropic with respect to the local 
direction of the magnetic field. 
The pressure tensor is
$\mathsf{P} \equiv p_\bot \mathsf{I} + (p_\parallel - p_\bot) \hat{\bb{b}} 
\hat{\bb{b}}$,
where
$\mathsf{I}$ stands for the $3 \times 3$ identity matrix. 
The symbols $\bot$ and $\parallel$ refer respectively to the directions 
perpendicular and parallel to the magnetic field $\bb{B}$, whose
direction is given by the unit vector $\hat{\bb{b}}\equiv \bb{B}/B = (b_x,
0, b_z)$. 
If the frequency of ion collisions $\nu_{ii}$ in the single ion species 
magneto-fluid is large compared to the rate of change $d/dt$ of all 
the fields involved, then the anisotropic part of the pressure tensor 
is small compared to its isotropic part 
$P\equiv 2p_\bot/3 + p_\parallel/3$, with 
$|p_\parallel - p_\bot| \ll P$.
This isotropic part of the pressure tensor is assumed to satisfy the equation of state for an ideal gas
\be
P=\frac{\rho k_B T}{\mu m_H}\,,\label{eq:EoS}
\en
where $k_B$ is the Boltzmann constant, $\mu$ is the mean molecular weight, and $m_H$ is the atomic mass unit. This equation along with Eqs.~(\ref{eq:rho})-(\ref {eq:S}) completes the specification of the dynamics of the unperturbed equilibrium configuration of the system under study. 

The anisotropic component of the pressure tensor in the momentum equation 
gives rise to Braginskii viscosity.  For small pressure anisotropy, this 
contribution is usually written as 
\be
p_\parallel- p_\bot = 3 \eta_0 \left(\hat{\bb{b}} \hat{\bb{b}} - \frac{1}{3}\mathsf{I} \right) : \bb{\nabla} \bb{v} \,, \label{eq:nu_braginskii}
\en
where $\eta_0$ is the largest of the coefficients in the viscous
stress tensor derived by \citet{1965RvPP....1..205B}[see also \citet{1985JGR....90.7620H}], and it is
related to the coefficient of kinematic viscosity via $\eta_0 = \rho
\nu_{\rm v}$.  
This is a good approximation provided that the pressure
anisotropy does not grow beyond $|p_\parallel-p_\bot|/P \simeq
\beta^{-1}$, where the plasma $\beta \equiv v_{\rm th}^2/v_{\rm A}^2$,
$v_{\rm th}\equiv (2P/\rho)^{1/2}$ is the thermal speed, and $v_{\rm
A}\equiv B/(4\pi\rho)^{1/2}$ is the Alfv{\'e}n speed.  
The effects of Braginskii viscosity in the linear dynamics of the
MTI and the HBI are explored in \cite{2011MNRAS.417..602K}.
Beyond this limit, various fast-growing, micro-scale plasma instabilities, such as
mirror and firehose (see \citealt{2005ApJ...629..139S,
2008PhRvL.100h1301S} and references therein) with growth rates $\gamma
\simeq k_\parallel v_{\rm th} |p_\parallel-p_\bot|/P$ can dominate the
plasma dynamics at very small scales. Thus, for
$|p_\parallel-p_\bot|/P \gtrsim \beta^{-1}$ the Braginskii-MHD
approximation embodied in Eqs.~(\ref{eq:rho})--(\ref{eq:S})
becomes ill-posed  and a mechanism to limit the pressure anisotropy 
must be implemented in numerical codes~\citep{2006ApJ...637..952S, 
2012ApJ...754..122K, 2012MNRAS.422..704P}.

Because the electron mean free path $({\lambda_{\rm mfp}})$ is large compared to its Larmor
radius, heat flows mainly along magnetic field lines.  This process is
modeled by the second term on the right-hand side of
Eq.~(\ref{eq:S}) via $\bb{Q}_{\rm s} \equiv-\chi\b(\b\bcdot
\bb{\nabla})T$, where $\chi$ is the thermal conductivity predominately
due to electrons with $\chi \approx 6 \times 10^{-7} T^{5/2} \,
\textrm{erg cm$^{-1}$ s$^{-1}$ K$^{-1}$}$ \citep{1962pfig.book.....S}.  In the equilibrium state, all the particles in
the plasma are assumed to be described by a Maxwellian distribution
with the same temperature, so that $p_\parallel\equiv p_\bot$
initially.  In general, the background heat flux does not vanish,
i.e., $\hat{\bb{b}}\bcdot \bb{\nabla} T \ne 0$, unless the magnetic
field and the background gradients are orthogonal.  The existence of a
well-defined steady state, i.e., $\bb{\nabla}\bcdot\bb{Q}_{\rm s}=0$,
demands that the background heat flux should be at most a linear
function of the distance along the direction of the magnetic field.

\begin{table}
 \begin{tabular}{lcc}
 \hline
 \hline
\textbf{Dimensional parameters}&\textbf{Symbols}&\textbf{Definitions}\\
 \hline
 \noalign{\vskip 0.05in}
Thermal speed of ions & $v_{\rm th}$ & $\sqrt{\frac{2P_i}{\rho_i}}$\\[1.2ex] 
Alfv\'en speed & $v_{\rm A}$ & $\frac{B}{\sqrt{4\pi\rho_i}}$\\[1.2ex]
Coefficient of thermal expansion & $\alpha$ & $-\f{1}{\rho}\left(\f{\pa\rho}{\pa T}\right)_P$\\[1.2ex]
Heat capacity at constant pressure & $c_p$ & $T\left(\frac{ds}{dT}\right)_P$\\[1.2ex]
Coefficient of kinematic viscosity & $\nu_{\rm v}$ & $\frac{\eta_0}{\rho}$\\[1.2ex]
Coefficient of thermal diffusion & $\lambda $ & $\f{\chi}{\rho c_p}$\\[1.2ex]
\hline
\textbf{Dimensionless parameters}&\textbf{Symbols}&\textbf{Definitions}\\
\hline
 \noalign{\vskip 0.05in}
Schwarzschild number & $ S $ & $\frac{g\alpha T d}{c_p\Delta T}-1$\\[1.2ex]
Rayleigh number & $R$ & $\f{\alpha(\Delta T) gd^3}{\lambda\nu_{\rm v}}$\\[1.2ex]
Chandrasekhar number & $Q$ & $\frac{B^2d^2}{4\pi\rho\nu_{\rm v} \eta} $\\ [1.2ex]
Prandtl number & ${\rm P_{\rm r}}$ & $\frac{\nu_{\rm v}}{\lambda} $\\[1.2ex]
Magnetic Prandtl number & ${\rm P_{\rm m}}$ & $\frac{\nu_{\rm v}}{\eta}$\\[1.2ex]
Plasma $\beta$ parameter & $\beta$ & $\frac{P_i}{B^2/8\pi}$\\[1.2ex]
Knudsen number & \textrm{Kn} & $\frac{\lambda_{\rm mfp}}{H}$\\[1.2ex]
\hline
\hline
\end{tabular}
\caption{List of parameters used in this letter.}
\end{table}

\subsection{Linear Equations for the Perturbations}
 
The equilibrium state $(\rho_*,P_*, T_*, v_*,B_*)$ 
is defined by the following relations
\be  
&& \bb{v}_*=0\,,\\
&&\f{dP_*}{dz}\approx\f{d(p_\bot)_*}{dz}=-\rho_* g\,,\label{eq:approxx}\\
&&\bb{B}_*=B_*\hat{\bb{b}}=B_*(\sin\phi\hat{\bb{x}}+\cos\phi\hat{\bb{z}})\,,\\
&&T_*= T_{\rm bottom}-\Delta T\,\left(\f{z}{d}\right)\,,
\en
where $\Delta T = T_{\rm bottom} - T_{\rm top}$,
and $d$ is the distance between the top and bottom boundaries.
As discussed above, the approximation invoked in Eq.~(\ref{eq:approxx}) 
reflects the fact that the pressure anisotropy is relatively weak.
We assume that the ion and electron pressures satisfy $P_i=P_e=P/2$ and
the ICM to be an ideal gas, and thus $\alpha T=1$, where 
$\alpha$ is the coefficient of thermal expansion. Note that this implies that the 
Schwarzschild number $S$, see Table 1, is constant even though $T$ changes with $z$.
Hereafter, we shall drop the asterisk subscripts denoting the equilibrium 
state as there will be no ambiguity. For the sake of convenience, Table 1 
provides a list of all the relevant parameters used in this letter.

Since the sound crossing time associated with the modes of interest is much shorter than the growth rate of the unstable modes of HBI and MTI, it is justified to work within the Boussinesq approximation
~\citep{2000ApJ...534..420B, 2001ApJ...562..909B, 2008ApJ...673..758Q}.
In this limit, Eq.~(\ref{eq:rho}) reduces to
\be
\bb{\nabla}\cdot\bb{v}=0\,.
\en
Thus, under the Boussinesq approximation, the velocity field perturbation satisfy $\bb{\nabla}\bcdot\delta\bb{v}=0$. Also, in this approximation the density variations can be ignored except when it appears multiplied with the external gravity term.

Together with the solenoidal character of the magnetic field fluctuations, 
$\bb{\nabla}\bcdot\delta\bb{B}=0$; the relation $\bb{\nabla}\bcdot\delta\bb{v}=0$ implies that it is only 
necessary to understand the dynamics of two independent components
for both the velocity and the magnetic field components. It is convenient
to use  as variables $\delta v_{z}, \delta \omega_{z}, \delta B_{z}$, and $\delta j_{z}$, where 
\be
&&\delta \omega_{z}\equiv  \partial_x \delta v_y - \partial_y \delta v_x\,,\\
&&\delta j_{z} \equiv   \partial_x \delta B_y - \partial_y \delta B_x\,,
\en
stand for the $z$-component of the fluctuations in the vorticity and the current density (times $4\pi$), respectively.

Taking the Laplacian of the $z$-component of the momentum Eq.~(\ref{eq:v})
and the $z$-component of its curl, we arrive to the equations of motion for
$\delta v_{z}$ and $\delta \omega_{z}$
\be
\label{eq:vz_fluc}
&&\pa_t\nabla^2\delta v_z=\alpha g(\pa_x^2+\pa_y^2)\delta T+ \nonumber \\
&&\frac{B}{4\pi\rho}(\sin\!\phi\, \, \pa_x+\cos\phi\,\pa_z)\nabla^2\delta B_z\nonumber\\
&&-3\nu_{\rm v} (\sin\!\phi\, \pa_x+\cos\phi\,\pa_z)^2[\cos\phi\,(\pa_x^2+\pa_y^2)-\sin\!\phi\, \pa_x\pa_z] \nonumber \\
&&\times(\sin\!\phi\, \delta v_x+\cos\phi\,\delta v_z)\,,
\en
\be
&&\pa_t\delta \omega_{z} = \frac{B}{4\pi\rho}(\sin\!\phi\, \pa_x+\cos\phi\,\pa_z)\delta j_{z} \nonumber \\
&&+3\nu_{\rm v} \sin\!\phi\, \pa_y(\sin\!\phi\, \pa_x+\cos\phi\,\pa_z)^2 (\sin\!\phi\, \delta v_x+\cos\phi\,\delta v_z)\,.\nonumber \\
\en
Following a similar procedure with the induction Eq.~(\ref{eq:b2}),
we obtain the equations for $\delta B_{z}$ and $\delta j_{z}$
\be
&&(\pa_t-\eta\nabla^2)\delta B_z=B(\sin\!\phi\, \pa_x+\cos\phi\,\pa_z)\delta v_z\,,
\en
\be
&&(\pa_t-\eta\nabla^2)\delta j_{z}=B(\sin\!\phi\, \pa_x+\cos\phi\,\pa_z)\delta \omega_{z}\,.
\en
We obtain the equation for the thermal fluctuations directly from 
Eq.~(\ref{eq:S}) as
\be
\label{eq:thermal_fluc}
&&\pa_t\delta T+\delta v_z\left(\frac{dT}{dz}+\frac{g\alpha T}{c_p}\right)=  \nonumber \\
&&\phantom{+}\lambda(\sin^{2}\!\phi\, \pa_x^2+2\sin\!\phi\,  \cos\phi\,\pa_x\pa_z+\cos^{2}\!\phi\, \pa_z^2)\delta T \nonumber\\
&&+\frac{\lambda}{B}\f{dT}{dz}(A_1\pa_y\delta B_y+A_2\pa_x\delta B_x+A_3\pa_z\delta B_z) \nonumber\\
&&+\frac{\lambda}{B}\f{dT}{dz}(A_4\pa_x\delta B_z+A_5\pa_z\delta B_x) \,,
\en
where we have defined  a number of functions in order to simplify the notation
\be
&&A_1\equiv \cos\!\phi\,, \\
&&A_2\equiv \cos\!\phi\,(\cos^{2}\!\phi-\sin^{2}\!\phi)\,, \\
&&A_3\equiv 2 \sin^{2}\!\phi\, \cos\!\phi\,, \\
&&A_4\equiv \sin\!\phi\, (\sin^{2}\!\phi-\cos^{2}\!\phi)\,, \\
&&A_5\equiv -2\sin\!\phi\, \cos^{2}\!\phi\,.
\en

\subsection{Dimensionless Variables}
It is convenient to use the characteristic scales in the problem in order to define 
a set of dimensionless coordinates according to
\be
&&\bb{x}'\equiv\bb{x}/d\,,\\
&&t'\equiv t\nu_{\rm v}/d^2\,.
\en
where $\nu_{\rm v}$ is the coefficient of kinematic viscosity.
This allows us to define a set of dimensionless functions for all the dynamical variables of interest,
i.e., 
$\delta v'_{z} \equiv \delta v_zd/\lambda$, 
$\delta \omega_{z}'\equiv \delta \omega_{z} d^2/\lambda$,
$\delta B'_z \equiv \delta B_z/B$,
$\delta j_{z}'\equiv \delta j_{z} d/B$, and 
$\delta \theta' \equiv \delta T/\Delta T$,
such that 
\be
\delta v'_{z}(\bb{x}', t')  &\equiv& \sum_{\bb{k}'} \, \hat{\delta v'_{z}}(z') \,\exp(i\bb{k}' \bcdot \bb{x}'+\sigma't')\,, \\
\delta \omega_{z}'(\bb{x}', t')  &\equiv& \sum_{\bb{k}'} \, \hat{\delta \omega'_{z}}(z') \,\exp(i\bb{k}' \bcdot \bb{x}'+\sigma't')\,, \\
\delta B'_z(\bb{x}', t') &\equiv& \sum_{\bb{k}'} \, \hat{\delta B'_{z}}(z') \,\exp(i\bb{k}' \bcdot \bb{x}'+\sigma't')\,, \\
\delta j_{z}'(\bb{x}', t')  &\equiv& \sum_{\bb{k}'} \, \hat{\delta j'_{z}}(z') \,\exp(i\bb{k}' \bcdot \bb{x}'+\sigma't')\,, \\
\delta \theta'(\bb{x}', t') &\equiv& \sum_{\bb{k}'} \, \hat{\delta \theta'}(z') \,\exp(i\bb{k}' \bcdot \bb{x}'+\sigma't') \,.
\en
Here, the hat-symbol ($\,\,\hat{}\,\,$) denotes the $z$-dependent Fourier transform amplitudes 
of the various functions 
involved, which are assumed to be periodic in the plane perpendicular to $z$.
The  dimensionless growth-rate (or frequency) $\sigma'$ characterizes the 
dynamics of the perturbation with dimensionless wave vector $\bb{k}'= (k'_x, k'_y, 0)$.

The linear nature of the
equations for the perturbations allows us to follow the dynamics of each mode 
independently.
For the sake of brevity, in what follows, unless otherwise specified, we omit all the 
primes labeling dimensionless coordinates, variables, and functions. We also 
omit the hat-symbol denoting the Fourier amplitude of a given mode.

Using the previous definitions,
the equations for the perturbations in equations (\ref{eq:vz_fluc})--(\ref{eq:thermal_fluc}), in dimensionless form, read

\be
\label{eq:pert_W}
&&\bigg[\sigma(\partial_z^2-k^2)-\frac{3}{k^2}(i \sin\!\phi\, k_x+ \cos\!\phi\, \partial_z)^2\times \nonumber \\
&&( \cos\!\phi\, k^2+i \sin\!\phi\, k_x\partial_z)^2\bigg]\delta v_z = -Rk^2\delta \theta\nonumber\\
&& + Q\f{{\rm P_{\rm r}}}{{\rm P_{\rm m}}}(i \sin\!\phi\, k_x+ \cos\!\phi\, \partial_z)(\partial_z^2-k^2)\delta B_{z}\nonumber\\
&&+\bigg[\frac{3i \sin\!\phi\, k_y}{k^2}(i \sin\!\phi\, k_x+ \cos\!\phi\, \partial_z)^2 \nonumber \\
&& ( \cos\!\phi\, k^2 +i \sin\!\phi\, k_x\partial_z)\bigg]\delta \omega_z\,,
\en
\be
\label{eq:pert_Z}
&&\left[\left(\frac{3\sin^{2}\!\phi\,k_y^2}{k^2}\right)(i\sin\!\phi\,k_x+\cos\!\phi\,\partial_z)^2\right]\delta \omega_z+\sigma\delta \omega_z= \nonumber \\
&&Q\f{{\rm P_{\rm r}}}{{\rm P_{\rm m}}}(i\sin\!\phi\,k_x+\cos\!\phi\,\partial_z)\delta j_{z} 
+\bigg[\f{3i\sin\!\phi\,k_y}{k^2}\times \nonumber \\ &&(i\sin\!\phi\,k_x+\cos\!\phi\,\partial_z)^2 \left(\cos\!\phi\,k^2+{i\sin\!\phi\,k_x}\partial_z\right)\bigg]\delta v_z\,,
\en
\be
&&(\partial_z^2-k^2-{\rm P_{\rm m}}\sigma)\delta B_{z}=-\f{{\rm P_{\rm m}}}{{\rm P_{\rm r}}}(i\sin\!\phi\,k_x+\cos\!\phi\,\partial_z)\delta v_z\,,\nonumber \\ \label{eq:pert_B}
\\
&&(\partial_z^2-k^2-{\rm P_{\rm m}}\sigma)\delta j_{z}=-\f{{\rm P_{\rm m}}}{{\rm P_{\rm r}}}(i\sin\!\phi\, k_x+\cos\!\phi\,\partial_z)\delta \omega_z\,,\nonumber \label{eq:pert_J}\\
\en
\be
&&(\cos^{2}\!\phi\,\partial_z^2+2i\sin\!\phi\,\cos\!\phi\,k_x\partial_z-\sin^{2}\!\phi\,k_x^2-{\rm P_{\rm r}} \sigma)\delta \theta \nonumber \\
&& =S\delta v_z+\f{1}{k^2}\left[({A_1}-{A_2})k_xk_y+i{A_5}k_y\partial_z\right]\delta j_{z}+ \nonumber\\
&& \bigg[-\f{A_1}{k^2}k_y^2\partial_z-\f{A_2}{k^2}k_x^2\partial_z+A_3\partial_z+iA_4k_x+i\f{A_5}{k^2}k_x\partial_z^2\bigg]\delta B_{z}\,.\nonumber \\
\label{eq:pert_T}
\en

In writing the preceding set of equations, we have also made use of the following relations:
\be
&&\delta v_x=(ik_x\pa_z\delta v_z+ik_y\delta \omega_{z})/k^2\,,\\
&& \delta v_y=(ik_y\pa_z\delta v_z-ik_x\delta \omega_{z})/k^2\,,\\
&&\delta B_x=(ik_x\pa_z\delta B_z+ik_y\delta j_{z})/k^2\,,\\ 
&& \delta B_y=(ik_y\pa_z\delta B_z-ik_x\delta j_{z})/k^2\,.
\en
\subsection{Boundary Conditions}

There are several sets of BCs that are commonly adopted in the framework of 
RBC. In what follows, we shall concern ourselves exclusively with the reflective, stress-free, and
perfectly conducting boundaries given by
\be
&&\delta v_z(0)=\delta v_z(1)=0\,,\label{W1}\\
&&\partial_z^2\delta v_z(0)=\partial_z^2\delta v_z(1)=0\,,\label{D2W}\\
&&\partial_z\delta \omega_z(0)=\partial_z\delta \omega_z(1)=0\,\label{DZ}, \\
&&\delta B_{z}(0)=\delta B_{z}(1)=0\,,\label{K}\\
&&\partial_z\delta j_{z}(0)=\partial_z\delta j_{z}(1)=0\,,\label{DX} \\
&&\delta \theta(0)=\delta \theta(1)=0\,\label{theta1} \,.
\en
Physically, Eq. (\ref{W1}) implies that the normal component of the velocity must
be zero on the boundary surfaces, whereas Eqs. (\ref{D2W}) and 
(\ref{DZ}) require stress-free surfaces and Eqs. (\ref{K})  and (\ref{DX}) 
imply perfectly conducting boundaries. Eq. (\ref{theta1}) fixes the boundary
surfaces to be at a constant temperature, 

For completeness, we state the type of BCs usually adopted in the literature 
related to the MTI and the HBI. The BC on the velocity is usually reflective, as 
embodied in Eqs.~(\ref{W1})--(\ref{DZ}), see, e.g., 
\cite{2011MNRAS.413.1295M, 2012ApJ...754..122K}. In these papers, the 
temperature is also fixed at the boundaries as in our Eq. (\ref{theta1}). 
The BCs employed in \cite{2012ApJ...754..122K} for the magnetic field are $\partial_z \delta 
B_x = \partial_z \delta B_y = 0$ (Eq. \ref{DX}) and $\delta B_z = 0$ (Eq. \ref{K}) 
when simulating an initially horizontal field  and $\delta B_x = \delta B_y = 0$ 
and  $\partial_z \delta B_z = 0$  when simulating an initially vertical field. 

The BCs that we use have the advantage of allowing us to derive analytic solutions.
In principle, we could have adopted BCs similar to the ones used in numerical 
simulations but this would in general require to solve the problem numerically, 
even in the linear regime. In passing, we may also mention that it is numerically straightforward 
to impose the BCs chosen in this letter making it possible to have future comparisons with numerical simulations.

\subsection{Relevant Characteristic ICM Values}

In the analysis that follows, it is of central importance to realize the extreme values that
some of the dimensionless parameters in Table 1 can reach under the conditions expected
in the ICM. For instance,  the dimensionless parameters $Q$ and ${\rm P_{\rm m}}$ have 
extremely large values. In order to set the scale, let us consider as an example
 \citep*{2002ARA&A..40..319C, 2006physrep...427..1}, e.g., $B\sim10^{-6}-10^{-7}$G, 
 $\rho\sim 10^{-27}-10^{-25}$gm cm$^{-3}$, $\eta=10-10^2$ cm$^2$ s$^{-1}$, $\nu_{\rm v}=10^{25}-10^{30}$ cm$^2$ s$^{-1}$, $T\sim10^{7}-10^{8}$K 
and radial length $d\sim10^{24}-10^{25}$cm. We thus find $Q\sim 10^{24}-10^{40}$ and ${\rm P_{\rm m}}\sim 10^{23}-10^{29}$.
This result suggests that it is justifiable to work in the limit in which $Q, {\rm P_{\rm m}} \rightarrow \infty$.
However, as we will show below, some of the results obtained in the stability analysis that follows from the
RB approach are sensitive to these limits being taken with the proper care.

\section{The Heat-Flux Driven Buoyancy Instability} \label{sec:HBI}

Let us first consider the case in which the magnetic field is
along the $z$-direction, i.e., $\phi=0$, which is known to be prone
to the HBI \citep{2008ApJ...673..758Q}. The RB formalism enables us to find the 
conditions for the existence of the HBI marginal state as follows.

In the state of marginal stability $(\sigma=0)$, the system of 
Eqs.~(\ref{eq:pert_W})--(\ref{eq:pert_Z}) reduces to
\be
&&3k^2\partial_z^2\delta v_z=-Q\f{{\rm P_{\rm r}}}{{\rm P_{\rm m}}}\partial_z(\partial_z^2-k^2)\delta B_{z}-Rk^2\delta \theta\,,\qquad\label{eq:Wm}\\
&&\partial_z^2\delta \theta=S\delta v_z-\partial_z\delta B_{z}\,,\label{eq:Tm}\\
&&(\partial_z^2-k^2)\delta B_{z}=-\f{{\rm P_{\rm m}}}{{\rm P_{\rm r}}}\partial_z\delta v_z\,,\label{eq:Km}\\
&&(\partial_z^2-k^2)\delta j_{z}=-\f{{\rm P_{\rm m}}}{{\rm P_{\rm r}}}\partial_z\delta \omega_z\,,\label{eq:Xm}\\
&&\partial_z\delta j_{z}=0\,,\label{eq:Zm}
\en
subject to the BCs,
$\delta v_z=[3k^2+Q]\partial_z^2\delta v_z=\partial_z^2\delta v_z=0, \partial_z\delta j_{z}=0, \partial_z\delta \omega_z=0, \delta B_{z}=0$ at $z=0,1$. 
Note that $\delta j_z$ and $\delta \omega_z$ have become decoupled from $\delta v_z$, $\delta B_z$, and $\delta\theta$.
From Eqs. (\ref{eq:Xm}) and (\ref{eq:Zm}) and the corresponding boundary 
conditions, it may be observed that, while the current density's vertical component 
vanishes identically and the vorticity's vertical component is independent of $z$  
at the marginal state we have
\be
\delta j_{z}={0}\,;\,\delta \omega_z={\rm constant}\,.
\en
The set of Eqs. (\ref{eq:Wm}), (\ref{eq:Tm}) and (\ref{eq:Km}) 
can be combined to give
\be
(3k^2\partial_z^4&-&SRk^2)(\partial_z^2-k^2)\delta v_z=\nonumber \\
&+&\left[R\f{{\rm P_{\rm m}}}{{\rm P_{\rm r}}}k^2+Q(\partial_z^2-k^2)\partial_z^2\right]\partial_z^2\delta v_z\,.\label{eq:Wmcfull}
\en
In the limit $Q,{\rm P_{\rm m}}\rightarrow\infty$ of interest, Eq. (\ref{eq:Wmcfull}) reduces to
 \be
 \left[R\f{{\rm P_{\rm m}}}{{\rm P_{\rm r}}}k^2+Q(\partial_z^2-k^2)\partial_z^2\right]\partial_z^2\delta v_z=0\,,\label{eq:Wmc}
 \en
 and also leads to the conclusion that at the boundaries, $z=0,1$,
\be
\partial_z^4(\partial_z^2-k^2)\delta v_z=0\,.
\en
The most general solution to (\ref{eq:Wmc}) has the form 
\be
\delta v_z(z)=C_0+C_1z&+&C_2\cos k_-z+C_3\sin k_-z \nonumber \\
&+&C_4\cosh k_+z+C_5\sinh k_+z\,,
\en
where $C_0,C_1,C_2,C_3,C_4$ and $C_5$ are constants of integration, and
\be
&& k_+\equiv \frac{k}{\sqrt{2}}\left[{\left(1+\f{4|R|{\rm P_{\rm m}}}{Q{\rm P_{\rm r}} k^2}\right)^{1/2}}+1\right]^{1/2}\,, \\
&& k_-\equiv \frac{k}{\sqrt{2}}\left[{\left(1+\f{4|R|{\rm P_{\rm m}}}{Q{\rm P_{\rm r}} k^2}\right)^{1/2}}-1\right]^{1/2}\,.
\en
Here, we have assumed that the upper boundary is hotter than the bottom one, as discussed below.
Therefore, the application of BCs allows us to conclude that $C_0=C_1=C_2=C_4=C_5=0$ and $k_{-}=n\pi$ ($n\in \mathbb{N}$), giving
\be
\delta v_z=C_3\sin n\pi z\,,
\en
along with
\be
R=-n^2\pi^2 \left[\f{n^2\pi^2+k^2}{k^2}\right]Q\f{{\rm P_{\rm r}}}{{\rm P_{\rm m}}}\,.\label{eq:mRscv}
\en
The marginal state can then exist only if $\Delta T$ is negative, i.e., 
if the upper boundary is hotter. 
Hence, by setting the temperature difference in the definition of the
Rayleigh number $R$ in Eq. (\ref{eq:mRscv}) to be $-\Delta T$, we obtain
\be
\f{-\alpha(\Delta T) gd^3}{\lambda\nu_{\rm v}}=-n^2\pi^2 \left[\f{n^2\pi^2+k^2}{k^2}\right]Q\f{{\rm P_{\rm r}}}{{\rm P_{\rm m}}}\,,
\en
and thus
\be
\f{d\ln T}{dz}=n^2\pi^2 \left[\f{n^2\pi^2+k^2}{k^2}\right]\left(\f{1}{\beta H}\right)\,,
\en
where we have set $d=H$, with $H$ the thermal-pressure scale height. 
Note that for a given $k$, the lowest value of ${d\ln T}/{dz}$ occurs when $n=1$ (\textit{lowest mode}) giving:
\be
\f{d\ln T}{dz}=\pi^2 \left[\f{\pi^2+k^2}{k^2}\right]\left(\f{1}{\beta H}\right)\,.
\en%
For all ${d\ln T}/{dz}$ smaller than this, all the perturbation with 
wavenumber $k$ are stable and they become unstable as this limit is overcome.
Since ${d\ln T}/{dz}$, for a given $n$, is a monotonically decreasing function of $k$, 
the minimum or \textit{critical} temperature gradient for the onset of the HBI occurs 
mathematically at  $k=k_c=\infty$. However, it is 
worth keeping in mind that in order for the fluid approach to remain valid, 
the wavenumber must satisfy $k \lesssim 2\pi/\lambda_{\rm mfp}$,
or in dimensionless numbers, $k\lesssim2\pi\textrm{Kn}^{-1}$. 
Therefore, using $k=2\pi\textrm{Kn}^{-1}$, the critical 
temperature gradient for the 
onset of the HBI is obtained as 
\be
\left.\f{d\ln T}{dz}\right|_c=\f{\pi^2(\textrm{Kn}^2+4)}{4\beta H}\approx\f{\pi^2}{\beta H}\,.
\label{eq:HBI_criterion_beta}
\en
This threshold for the temperature gradient takes into
account the effect of magnetic tension induced by a finite value
of the plasma $\beta$ parameter, which has been usually 
ignored when deriving the stability criterion for the HBI.
In the limit of $\beta\rightarrow\infty$, Eq. 
(\ref{eq:HBI_criterion_beta}) recovers the usual criterion for the HBI
\citep{2008ApJ...673..758Q}.

In general, one can use variational principles to investigate the presence of oscillatory marginal states
and the validity of the principle of exchange of instabilities \citep{1981HH.book.....}.
However, in what follows we show that the lowest
mode always appears as a stationary state for the HBI.
For this purpose we set $\sigma=iq$, $q$ being real, and rewrite the relevant Eqs.~(\ref{eq:pert_W}), (\ref{eq:pert_B}), and (\ref{eq:pert_T})
\begin{align}
&\left[iq(\partial_z^2-k^2)-3k^2\partial_z^2\right]\delta v_z=Q\f{{\rm P_{\rm r}}}{{\rm P_{\rm m}}}\partial_z(\partial_z^2-k^2)\delta B_{z} -Rk^2\delta \theta \,,\label{eq:Wmo}\\
&(\partial_z^2-k^2-i{\rm P_{\rm m}}q)\delta B_{z}=-\f{{\rm P_{\rm m}}}{{\rm P_{\rm r}}}\partial_z\delta v_z\,.\label{eq:Kmo}\\
&(\partial_z^2-i{\rm P_{\rm r}} q)\delta \theta=S\delta v_z-\partial_z\delta B_{z}\,,\label{eq:Tmo}
\end{align}
In the limit $Q,{\rm P_{\rm m}}\rightarrow\infty$ of interest,
it is trivial to arrive at the following using the real and imaginary
parts of the previous three equations:
\be
&& q^2{\rm P_{\rm m}}[(\partial_z^2-k^2)\partial_z^2+3{\rm P_{\rm r}} k^2\partial_z^2]\delta v_z=\nonumber\\
&&-Q(\partial_z^2-k^2)\partial_z^4\delta v_z-R\f{{\rm P_{\rm m}}}{{\rm P_{\rm r}}}k^2\partial_z^2\delta v_z\,,\label{eq:realo}\qquad\\
&&
-q{\rm P_{\rm m}}[-3k^2\partial_z^4+q^2{\rm P_{\rm r}}(\partial_z^2-k^2)+RSk^2]\delta v_z=
\nonumber\\
&& q{\rm P_{\rm r}} Q(\partial_z^2-k^2)\partial_z^2\delta v_z\,.\label{eq:imago}
\en
Using $\delta v_z=C_3\sin\pi z$ as the lowest mode for a top-hot-plate configuration and  
making use of Eqs. (\ref{eq:realo}) and (\ref{eq:imago}),
one can arrive at the following relations
\be
&&q^2{\rm P_{\rm m}}\left[(\pi^2+k^2)-3{\rm P_{\rm r}} k^2\right]=Q(\pi^2+k^2)\pi^2+R\f{{\rm P_{\rm m}}}{{\rm P_{\rm r}}}k^2\,,\nonumber
\label{eq:realop}\\
\en
and
\be 
q{\rm P_{\rm m}}[3k^2\pi^4+q^2{\rm P_{\rm r}}(\pi^2+k^2)-RSk^2]= \nonumber \\
-q{\rm P_{\rm r}} Q(\pi^2+k^2)\pi^2 \,.
\label{eq:imagop}
\en

Multiplying (\ref{eq:realop}) by $q{\rm P_{\rm r}}$ and adding with (\ref{eq:imagop}), 
we conclude that either $q=0$ or
\be
3q^2{\rm P_{\rm r}}^2=R(S-1)-3\pi^4\,.\label{eq:oms_hbi}
\en
For typical values of the parameters in a galaxy cluster, Eq. (~\ref{eq:oms_hbi}) can be satisfied for all \bb{k} and for any real q. Therefore the lowest unstable HBI mode can set in as an oscillatory marginal state.

\section{The Magnetothermal Instability} \label{sec:MTI}

When the magnetic field is aligned with the $x$-direction, and thus it is 
perpendicular to the direction of gravity, i.e., 
$\phi=\pi/2$, the plasma may be subject to the MTI \citep{2001ApJ...562..909B}.
At marginal stability, we can formulate the problem 
by setting $\sigma=0$ in Eqs. (\ref{eq:pert_W})--(\ref{eq:pert_T}), which become
\begin{align}
&-\f{3}{k^2}k_x^4\partial_z^2\delta v_z=-Rk^2\delta \theta+
iQ\f{{\rm P_{\rm r}}}{{\rm P_{\rm m}}}k_x(\partial_z^2-k^2)\delta B_{z}+\frac{3k_x^3k_y}{k^2}\partial_z\delta \omega_z\,,
\label{eq:Wmsh}\\
&-\frac{3k_x^2k_y^2}{k^2}\delta \omega_z=\f{3k_x^3k_y}{k^2}\partial_z\delta v_z+iQ\f{{\rm P_{\rm r}}}{{\rm P_{\rm m}}}k_x\delta j_{z}\,,\label{eq:Zmsh} \\
&(\partial_z^2-k^2)\delta B_{z}=-i\f{{\rm P_{\rm m}}}{{\rm P_{\rm r}}}k_x\delta v_z\,,\label{eq:Kmsh}\\
&(\partial_z^2-k^2)\delta j_{z}=-i\f{{\rm P_{\rm m}}}{{\rm P_{\rm r}}}k_x\delta \omega_z\,,\label{eq:Xmsh}\\
&-k_x^2\delta \theta=S\delta v_z+ik_x\delta B_{z}\,,\label{eq:Tmsh}
\end{align}
subject to the BCs, at $z=0,1$, 
\be
&&\delta \theta=0\,;\,\delta v_z=\partial_z^2\delta v_z=0\, ;\nonumber \\ 
&& \partial_z\delta j_{z}=0\,;\,\partial_z\delta \omega_z=0\,;\,  \delta B_{z}=0\,.
\en 
One can see that the BC for the temperature $\delta \theta$ is 
trivially satisfied due to the BCs on $\delta v_z$ and $\delta B_z$. 
Combining Eqs. (\ref{eq:Wmsh})--(\ref{eq:Zmsh}), we get the following differential equation for $\delta v_z$ 
\begin{align}
&Q\left[RSk^2k_x^2(\partial_z^2-k^2)+\f{3}{k^2}k_x^8\partial_z^2(\partial_z^2-k^2)+\f{3}{k^2}k_x^6k_y^2(\partial_z^2-k^2)^2\right]\delta v_z\nonumber\\
&+Q^2k_x^6(\partial_z^2-k^2)\delta v_z+QR\f{{\rm P_{\rm m}}}{{\rm P_{\rm r}}}k^2k^4_x\delta v_z+3RSk_x^2k_y^2(\partial_z^2-k^2)^2\delta v_z\nonumber\\
&+3Rk_x^4k_y^2\f{{\rm P_{\rm m}}}{{\rm P_{\rm r}}}(\partial_z^2-k^2)\delta v_z=0\,.\label{eq:disp_mti}
\end{align}
We focus on the limit of interest, i.e., 
$Q,{\rm P_{\rm m}}\rightarrow\infty$, 
which in this case allows us to write the following single differential equation for $\delta v_z$,
where it is only necessary to retain terms up to order 
$Q^2\,,{\rm P_{\rm m}}^2$ and $Q{\rm P_{\rm m}}$ have been kept:
\be
Q^2k_x^6(\partial_z^2-k^2)\delta v_z+QR\f{{\rm P_{\rm m}}}{{\rm P_{\rm r}}}k^2k^4_x\delta v_z=0\,.\label{eq:Wmc2}
\en
Taking the successive even $z-$derivatives of this equation and evaluating the
results at $z=0,1$ we obtain
\be
\partial_z^{2m}\delta v_z=0\,\quad \textrm{with}\,\quad m\in\{0,1,2,...\}\,.
\en
This means that the appropriate solution for the lowest mode is 
\be
\delta v_z=A\sin\pi z\,, 
\en
where $A$ is a constant. Therefore, Eq. (\ref{eq:Wmc2}) implies:
\be
R=\f{k_x^2}{k^2}(\pi^2+k^2)Q\f{{\rm P_{\rm r}}}{{\rm P_{\rm m}}}\,,\label{eq:RQrel}
\en
where $k_x\ne0$.
The marginal state can exist only if $\Delta T$ is positive, i.e., if the bottom boundary is hotter. 
Note that for fixed $k_x$, the Rayleigh number $R$ monotonically decreases as 
$k_y$ increases; and, for fixed $k_y$, $R$ monotonically increases with  $k_x$.
Also, recall that in order for the local analysis in $(x,y)$ 
 to be valid within the fluid approach we should have $H^{-1}< k_x, k_y < \lambda^{-1}_\textrm{mfp}$
Therefore, we note that, while the minimum possible value of $k_x$ is $(k_x)_{\rm min}\gtrsim2\pi$, the maximum value of $k_y$ is $(k_y)_{\rm max}\lesssim2\pi/\textrm{Kn}$ and this combination $(k_x,k_y)$ corresponds to the lowest unstable mode.

The minimum value for the temperature gradient $-d\ln T/dz$ for the onset of MTI 
modes with $k_x, k_y\ne0$, is obtained from 
\be
\left.-\f{d\ln T}{dz}\right|_c&=&\min\left[\f{k_x^2(\pi^2+k_x^2+k_y^2)}{k_x^2+k_y^2}\right]\left(\f{1}{\beta H}\right)\,,
\label{eq:MTI_criterion_beta}
\en
where the minimum value
\be
&&\min\left[\f{k_x^2(\pi^2+k_x^2+k_y^2)}{k_x^2+k_y^2}\right]
\equiv \nonumber \\ && \qquad \qquad \frac{{(k_x)_{\rm min}^2[\pi^2+(k_x)_{\rm min}^2+(k_y)_{\rm max}^2}]}{{(k_x)_{\rm min}^2+(k_y)_{\rm max}^2}}\,,
\en
depends on the particular mode under consideration.
This threshold for the temperature gradient takes into
account the effect of magnetic tension induced by a finite value
of the plasma $\beta$ parameter, which has been usually 
ignored when deriving the stability criterion for the MTI.
In the limit of $\beta\rightarrow\infty$, Eq. 
(\ref{eq:MTI_criterion_beta}) recovers the usual criterion for the MTI
\citep{2001ApJ...562..909B}.

We now show that the lowest mode doesn't 
set in as an oscillatory marginal stability state. In order to achieve this,
let us define $\sigma=iq$, with $q$ real and write the relevant version of Eqs.~(\ref{eq:pert_W})--(\ref{eq:pert_T}) as
\be
&&\left[iq(\partial_z^2-k^2)-\f{3}{k^2}k_x^4\partial_z^2\right]\delta v_z=iQ\f{{\rm P_{\rm r}}}{{\rm P_{\rm m}}}k_x(\partial_z^2-k^2)\delta B_{z}\nonumber\\
&&-Rk^2\delta \theta +\frac{3k_x^3k_y}{k^2}\partial_z\delta \omega_z\,,\qquad\\
&&\left(-\frac{3k_x^2k_y^2}{k^2}-iq\right)\delta \omega_z=\f{3k_x^3k_y}{k^2}\partial_z\delta v_z+iQ\f{{\rm P_{\rm r}}}{{\rm P_{\rm m}}}k_x\delta j_{z}\,.\nonumber \\
&&(\partial_z^2-k^2-i{\rm P_{\rm m}}q)\delta B_{z}=-i\f{{\rm P_{\rm m}}}{{\rm P_{\rm r}}}k_x\delta v_z\,,\\
&&(\partial_z^2-k^2-i{\rm P_{\rm m}}q)\delta j_{z}=-i\f{{\rm P_{\rm m}}}{{\rm P_{\rm r}}}k_x\delta \omega_z\,,\\
&&(-k_x^2-i{\rm P_{\rm r}} q)\delta \theta=S\delta v_z+ik_x\delta B_{z}\,,
\en
In the limit $Q,{\rm P_{\rm m}}\rightarrow\infty$, one can as before argue in favour of using $\delta v_z=A\sin\pi z$ as 
the lowest mode. 
Using the immediately preceding five equations, 
it is easy to arrive at the following real and 
imaginary parts of a complex equation, leading to
\be
&&q^4\left[(\pi^2+k^2)\left(k_x^2+\f{k_x^2k_y^2}{k^2}{\rm P_{\rm r}}\right)-\f{3k_x^4}{k^2}{\rm P_{\rm r}}\pi^2\right] \nonumber \\
&&+ q^2\left[\f{Rk_x^2k^2}{{\rm P_{\rm r}}}+3RSk_x^2k_y^2\right] \nonumber \\
&&-q^2\left[\f{Qk_x^2}{{\rm P_{\rm m}}}\left(k_x^2-\f{3k_x^2k_y^2}{k^2}{\rm P_{\rm r}}\right)(\pi^2+k^2)\right]=0\,,
\en
\be
&&q^5{\rm P_{\rm r}}(\pi^2+k^2)
+q^3\left[\f{3k_x^6}{k^2}\pi^2+\f{3k_x^4k_y^2}{k^2}(\pi^2+k^2)\right]
\nonumber\\&&
-q^3\left[RSk^2+\f{Q{\rm P_{\rm r}}}{{\rm P_{\rm m}}}k_x^2(\pi^2+k^2)\right]
\nonumber\\&&
-q\left[\left(\f{Q}{{\rm P_{\rm m}}}\right)\f{3k_x^6k_y^2}{k^2}(\pi^2+k^2)-\f{3k_x^4k_y^2}{{\rm P_{\rm r}}}R\right]=0\,.\qquad\quad
\en
In the light of Eq.~(\ref{eq:RQrel}),
these two equations simplify respectively to 
\begin{align}
&q^4\left[(\pi^2+k^2)\left(k_x^2+\f{k_x^2k_y^2}{k^2}{\rm P_{\rm r}}\right)-\f{3k_x^4}{k^2}{\rm P_{\rm r}}\pi^2\right]+ q^2\left[3R(S+1)k_x^2k_y^2\right]=0\,,\qquad\qquad\\
&q^5\left[{\rm P_{\rm r}}(\pi^2+k^2)\right]+q^3\left[\f{3k_x^6}{k^2}\pi^2+\f{3k_x^4k_y^2}{k^2}(\pi^2+k^2)-R(S+1)k^2\right] =0 \,.
\end{align}
 Since $\Delta T>0$ in this case, all the terms inside the square brackets are positive definite allowing us to conclude that for the existence of an oscillatory marginal state $q\ne0$, we must have
 \begin{align}
& \left|
   \begin{array}{ll} 
  m_{11} & m_{12} \\ 
  m_{21} & m_{22} \\ 
  \end{array}
   \right|=0\,.\label{eq:detcond}
 \end{align} 
where the elements in the determinant are 
\begin{align}
&   m_{11} = (\pi^2+k^2)\left(k_x^2+\f{k_x^2k_y^2}{k^2}{\rm P_{\rm r}}\right)-\f{3k_x^4}{k^2}{\rm P_{\rm r}}\pi^2 \,, \\
&   m_{12} = 3R(S+1)k_x^2k_y^2 \,, \\
&   m_{21} =   {\rm P_{\rm r}}(\pi^2+k^2) \,, \\
&   m_{22} = \f{3k_x^6}{k^2}\pi^2 +\f{3k_x^4k_y^2(\pi^2+k^2)}{k^2}- R(S+1)k^2 \,.
\end{align}
While this condition, in principle, allows for the existence of a marginal oscillatory stability state for MTI, it is not easy to explicitly specify analytically the values of $k_x$~and~$k_y$ for which this happens. 
Thus, without solving explicitly this equation in terms of $(k_x,k_y)$, it is not possible for us to rule out the onset of the MTI as an oscillatory mode. 
%
%


\subsection{Retrieving Schwarzschild Criterion}
Before we end this section, it is instructive to note how the limits 
$Q,{\rm P_{\rm m}}\rightarrow\infty$ play an important role to bring about the correct 
criterion for the onset of the MTI.

Let us consider modes with $k_x\rightarrow0$, \textit{without necessarily}
imposing the limits $Q,{\rm P_{\rm m}}\rightarrow\infty$. 
In this case, using (\ref{eq:Wmsh})-(\ref{eq:Zmsh}), we arrive at
\be
&&\f{3k_x^8}{k^2}\mathfrak{D}\partial_z^2(\partial_z^2-k^2)\delta v_z+\f{9k_x^8k_y^2}{a^4}\partial_z^2(\partial_z^2-k^2)^2\delta v_z=  \\ 
&&Qk_x^6(\partial_z^2-k^2)\mathfrak{D}\delta v_z+RSk^2k_x^2(\partial_z^2-k^2)\mathfrak{D}\delta v_z+\nonumber\\
&&R\f{{\rm P_{\rm m}}}{{\rm P_{\rm r}}}k^2k_x^4\mathfrak{D}\delta v_z\,, \nonumber
\en
where we have introduced, $\mathfrak{D}\equiv Q+3(k_y^2/k^2)(\partial_z^2-k^2)$. Hence taking the limit $k_x\rightarrow0$ yields
\be
RSk^2(\partial_z^2-k^2)\left[Q+3\f{k_y^2}{k^2}(\partial_z^2-k^2)\right]\delta v_z=0\,.
\en
Using $\delta v_z=A\sin\pi z$ as the lowest mode, the condition for marginally stable state is $RS=0$, i.e.,
\be
\Delta T \, \left[\frac{g\alpha T d}{c_p(\Delta T)}-1\right]=0\,,
\en
and thus
\be
-\f{dT}{dz}=\frac{g\alpha T }{c_p}\,.
\label{eq:Schwarzschild}
\en
This condition, which is independent of the choice of a specific mode for $\delta v_z$, 
is just the condition for the marginal state corresponding to the Schwarzschild 
instability. 
Perhaps a more direct way to  arrive to this condition is to set $k_x=0$ in Eq.
(\ref{eq:Tmsh}) and assume that $\delta v_z\ne0$, which leads to the conclusion 
that the Schwarzschild number is $S=0$, implying that  
$-{dT}/{dz}={g\alpha T }/{c_p}$, as stated in Eq. 
(\ref{eq:Schwarzschild}).

It is not difficult to understand the physics that allow us to retrieve the Schwarzschild instability criterion in the limit $k_x\rightarrow0$. In an unmagnetized stratified atmosphere, a fluid element that is adiabatically displaced upwards returns to its initial position if the entropy gradient is positive. This condition is known as Schwarzschild criterion. Now, in the case of the MTI, an upwardly displaced fluid element carries the magnetic field along while retaining its temperature unchanged because the heat quickly flows along the magnetic field lines. This mechanism leads to the MTI. When a perturbation with $k_x\rightarrow 0$ is considered, we could envision the associated mode to have an infintely long wavelength and thus a fluid parcel displaced upwards is not connected via magnetic field lines to its initial position (i.e., the horizontal layer of atmosphere where it was initially in). Therefore, heat is unable to flow into the displaced layer and one obtains back the Schwarzschild criterion for instability. In this context, it may be noted how setting $k_x=0$ prevents the magnetic field and the temperature perturbations in Eq. (\ref{eq:Tmsh}) from coupling to the velocity perturbation.

\section{Summary and Discussion} 
\label{sec:DaC}

In this letter, we have applied the formalism employed in RBC to
study the MTI and the HBI. This approach goes beyond the standard
linear mode analysis that has been carried out (but see
\citealt{2012MNRAS.423.1964L} for an exception) by considering
explicit boundary conditions. This enabled us to address in a natural
way, some aspects of the linear dynamics of these instabilities
that have not been previously addressed.

In particular, we have derived the conditions for the onset of the
instabilities retaining the effects of magnetic tension, as embodied by
a finite plasma beta parameter, and Braginskii viscosity.
The latter is known to have a stabilizing effect on the 
high-$k$ end of the spectrum of unstable modes \citep{2011MNRAS.417..602K}.
We found, however, that (to linear order) 
Braginskii viscosity does not play an explicit role in the criterion for 
the onset of either the HBI or the MTI, see Eqs. (\ref{eq:HBI_criterion_beta}) and (\ref{eq:MTI_criterion_beta}). 
We have found expressions
for the Rayleigh number in terms of the wave vector
$\bb{k}$ of a given mode. In the case of the HBI, the Rayleigh number
is found to be a monotonically decreasing function of the mode
wavenumber, $k$. This implies that, for a given temperature gradient,
the modes that go unstable first are those with largest values of $k$,
i.e., those with smallest wavelength. In the case of the MTI, the
dependence of the Rayleigh number on the dimensionless wave vector
$\bb{k}$ is more subtle, as it depends on both its magnitude and
direction. The MTI modes that go unstable first are those with long
wavenumber while maintaining $k_{||}\rightarrow0$, which is a
restriction that does not apply to HBI. For the HBI, the 
mode that goes unstable first does so in a  non-oscillatory fashion, 
whereas in the MTI, an oscillatory marginal stable state is, in principle, possible.

We have found that the HBI is regularized (at high wavenumbers) by magnetic tension,
but such regularization is not present for the MTI.  For collisional plasmas, there is a high-$k$ regularization by isotropic viscosity and conductivity.  In the weakly-collisional regime that concerns us here, the transport is dominant along magnetic field lines. Nevertheless, there is still some, albeit small, isotropic diffusion.  In order to find a critical Rayleigh number for the MTI, as needed to perform a weakly nonlinear analysis, it might be useful to include this isotropic contribution. While it is certainly possible to include this effect in the equations, this would render the analytical treatment that we have presented significantly more challenging. It is thus pragmatic to defer this calculation to future work while allowing the present work to focus solely on the nontrivial effects of anisotropic heat conduction.

Before we conclude, we comment on the choice of the BCs. For the sake
of analytical simplicity, we have chosen to work with conducting
stress-free BC. Other possibilities include non-conducting
stress-free, conducting rigid, non-conducting rigid, etc. For some
of these BCs, it is not possible to solve the problem solely on
analytical grounds, and numerical techniques are necessary even in the
linear regime. The BC we have employed resemble those usually employed
in numerical simulations. The specific choice of boundary conditions
is unlikely to have a dramatic impact in 
the stability criterion within the bulk of the plasma.
Nevertheless, it should be kept in mind that these
could have an impact on the specific expression for the stability criteria.

Some of these results could have been obtained by other means, for
example by retaining the effects of magnetic tension and solving
analytically the associated dispersion relations. However, the
formalism we have outlined could become even more advantageous as the
dispersion relations dictating the linear dynamics become more
involved. It is worth noticing that the formalism can be generalized
to address more realistic physical settings, for example including the
 effects of cosmic-rays \citep{2006ApJ...642..140C}, 
 rotation \citep{2014APJ...792..1}, and radiative cooling
 \citep{2010ApJ...720L..97B, 2012MNRAS.423.1964L}, or composition 
 gradients in the ICM \citep{2013ApJ...764..13, 2015ApJ...813...22B}

One advantage of having laid out the RBC formalism is that this
provides the grounds for future work on weakly-non linear analysis.
This type of analysis has proven to be advantageous in delivering
further analytical insights into the bifurcation scenarios and the
routes to chaotic (turbulent) states of the systems under study 
\citep{1987CCF.book.....,1999WS.book.....}.

\section*{Acknowledgments}
We are thankful to Thomas Berlok and Mahendra K. Verma
for useful discussions. The research leading to these results has
received funding from the European Research Council under the European
Union's Seventh Framework Programme (FP/2007-2013) under ERC grant
agreement 306614 (MEP). MEP also acknowledges support from the Young
Investigator Programme of the Villum Foundation (VKR022591). SC gratefully
acknowledges the financial support from INSPIRE faculty award (DST/INSPIRE/04/2013/000365) conferred by the 
Indian National Science Academy (INSA) and the Department of Science and Technology (DST), India.
\section*{References}
\bibliographystyle{elsarticle-harv} 
\bibliography{Gupta_etal}
\end{document}